\documentclass[twocolumn]{aastex701}  
\usepackage{graphicx}
\usepackage{xcolor}
\usepackage{amsmath}
\usepackage{txfonts}
\usepackage{tikz}
\usepackage{caption}
\usepackage{float}
\usepackage{subcaption}
\usepackage{comment}
\begin{document}
\title{Planets in Pulsar Winds}


\author{T. Kaister}
\affiliation{Nicolaus Copernicus Astronomical Center, Polish Academy of Sciences, Bartycka 18, 00-716, Warsaw, Poland}
\email{tkaister@camk.edu.pl}

\author{S. Andr\'{e}s Joya M\'{e}ndez}
\affiliation{Physikalisches Institut, University of Bern, Sidlerstrasse 5, 3012 Bern, Switzerland}
\email{sergio.joya@students.unibe.ch}

\author{P. Marmat}
\affiliation{Janusz Gil Institute of Astronomy, University of Zielona G\'{o}ra, ul. Prof. Z. Szafrana 2, 65-516 Zielona G\'{o}ra, Poland}
\email{pmaramat@ph.iitr.ac.in}

\author{M. \v{C}emelji\'{c}}
\affiliation{Nicolaus Copernicus Astronomical Center, Polish Academy of Sciences, Bartycka 18, 00-716, Warsaw, Poland}
\affiliation{Nicolaus Copernicus Superior School, College of Astronomy and Natural Sciences, ul. Nowogrodzka 47a, 00-695, Warsaw, Poland}
\affiliation{Research Centre for Computational Physics and Data Processing, Institute of Physics, Silesian University in Opava, Bezru\v{c}ovo n\'{a}m.~13, CZ-746\,01 Opava, Czech Republic}
\email{miki@camk.edu.pl}

\author{M. Velli}
\affiliation{Department of Earth, Planetary, and Space Sciences, University of California--Los Angeles, Los Angeles, CA 90056, USA}
\email{mvelli@ucla.edu}

\author{J. Varela}
\affiliation{Institute for Fusion Studies, Department of Physics, University of Texas at Austin, Austin, Texas 78712, USA}
\email{Jacobo.Rodriguez@austin.utexas.edu}

\author{M. Falanga}
\altaffiliation{Deceased: 2025 March 6.}
\affiliation{Physikalisches Institut, University of Bern, Sidlerstrasse 5, 3012 Bern, Switzerland}
\affiliation{International Space Science Institute (ISSI), Hallerstrasse 6, 3012 Bern, Switzerland}
\email{issi@issibern.ch}
 
\begin{abstract}
Planets around pulsars were the first discovered exoplanets, found thanks to the extremely precise pulsar timing. Here we suggest that they could also be found through the radio emission generated by the pulsar-planet magnetospheric interaction. We present the results of special relativistic numerical simulations of planets in a pulsar wind of velocity $v=0.985~c$, corresponding to a Lorentz factor $\gamma=5.795$. Planets, modeled as a perfectly conducting solid surface in an external magnetic field originating from the pulsar, produce radio emission in the extended magnetic structure on the planet's nightside. We find that the planet around a known pulsar, PSR J0636+5129 b, could be detected via its radio emission. We outline the observational prospects for such objects.
\end{abstract}


%
\section{Introduction}
Planets and pulsars seem an unlikely pairing. Pulsars are among the most extreme objects in the universe — rapidly rotating neutron stars that bathe their surroundings in relativistic winds and high-energy radiation. However, it was precisely around such a hostile object that the first confirmed exoplanets were discovered \citep{Wolszczan1992}\footnote{Planets around pulsars were first suggested in \cite{DemPro79} around the bright pulsar PSR B0329+54; they were not later confirmed, but this system remains a tentative case \citep{StaRod17}.}. Since that discovery, only 13 pulsar planets have been confirmed; three orders of magnitude fewer than those found around main-sequence stars \citep{NASAExoplanetArchivePS}. Recent sky surveys such as \cite{Ni_u_2022} and \cite{Behrens_2020} have not found any new systems despite their high sensitivity. However, new objects are still being found. The MeerKAT telescope recently discovered a companion to pulsar M62H with a median mass of $\approx 3M_J$ \citep{vleeschower2024discoveriestimingpulsarsm62}; whether this object is a true planet or a degenerate stellar remnant is still debated. This scarcity raises a fundamental question: are pulsar planets intrinsically rare, or are our detection methods simply not sensitive enough to find them?

To date, all confirmed pulsar planets have been discovered using pulsar timing: a technique that leverages the extraordinary regularity of neutron star radio pulses, which can be perturbed by the gravitational tug of orbiting planets. This method, while highly precise, does not offer insight into the magnetospheric environment or physical conditions at the planet. To further probe the planetary conditions in this extreme environment, alternative methods are needed.

In this work, we propose a novel method for identifying pulsar planets via their radio emissions, which are generated by the interaction between a planet and the pulsar wind. Following the theoretical framework developed by \cite{refId1, refId2}, we expect strong Alfvén wings, analogous to the Io–Jupiter interaction \citep{https://doi.org/10.1029/JA085iA03p01171} but in a relativistic regime. The Alfvén wings are predicted to be powerful radio emitters, with the emitted power scaling with the local magnetic Poynting flux \citep{Zarka07}. Furthermore, the emission is highly collimated by relativistic aberration \citep{refId0}, making it in principle detectable at interstellar distances. Analyzing the predicted radio emission would allow for further study of the pulsar wind environment: Alfvénic structures in pulsar winds have been shown to be highly sensitive to plasma composition, geometry, and pair multiplicity \citep{singh2025selfconsistentmodelkineticalfven}, suggesting that observed emission spectra could in principle constrain the ionic content of the wind. 

The aim of this work is to investigate the feasibility of detecting radio emission from nearby pulsar planets. Using 3D relativistic MHD simulations implemented in PLUTO \citep{Mignone_2007}, we model the planet as a perfectly conducting or ferromagnetic surface, exposed to an ultra-relativistic pulsar wind with a Lorentz factor $\gamma = 5.798$. This extends the study by \citet[hereafter Paper I]{mishra2023aurorasplanetspulsars}, which examined a similar setup at $\gamma = 2.0$. We advance the state of the art in three directions. First, we increase the Lorentz factor of the simulated pulsar wind from $\gamma = 2.0$ (Paper I) to $\gamma = 5.8$. Real pulsar winds are estimated to reach $\gamma \approx 10^4
-10^5$\citep{2012Natur.482..507A}, placing our simulations in an intermediate but more physically realistic regime than Paper I. For simulations, $\gamma > 10.0$ is rarely used, as at that point relativistic MHD codes reach their computational limits and semi-analytical methods would be needed. Second, we include emission from the planet nightside structure, which, as we show, approximately doubles the total radiated power. Third, we go beyond qualitative, extrapolated detectability arguments from Paper I and compute explicit flux density estimates for known and hypothetical systems, benchmarked against the sensitivities of LOFAR, MeerKAT, and the forthcoming SKA.
We find that for the known pulsar planet PSR J0636+5129 b, the predicted emission is detectable with currently operating telescopes, making it a suitable target for an observational follow-up.

In Section~\ref{Pluto} we present the details of our simulation setup. The methods used to analyze the simulations are described in Sections~\ref{comput} and \ref{shape}, with applications to specific exoplanets detailed in Section~\ref{planets}. An observational perspective is provided in Section~\ref{obs}, and a brief recapitulation of the main points is presented in the Conclusions.

\section{Setup in PLUTO code}\label{Pluto}
\begin{figure}[h]
    \includegraphics[width=\columnwidth]{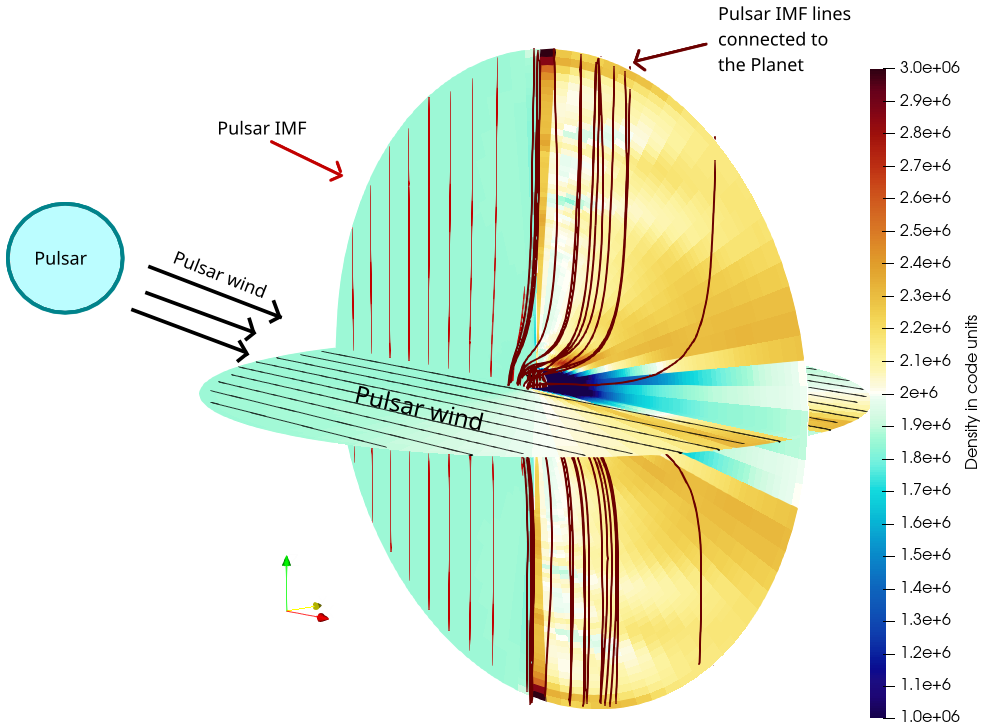}
    \caption{Schematic of the setup. Far from the pulsar, its magnetic field is assumed to be nearly homogeneous; at the planet's location, it is perpendicular to the pulsar wind. Under these external boundary conditions, the wind pressure bends the field around the planet.}
    \label{setup}
\end{figure}
We perform three-dimensional (3D) numerical simulations of the magnetospheric interaction between a non-magnetized conductive planet and a highly relativistic pulsar wind using the PLUTO code \citep{Mignone_2007}. Our model is shown in Fig.~\ref{setup}. At the inner boundary of the computational domain, which represents the planetary surface, we assume that the radial magnetic field vanishes and that all non-radial electric fields also vanish, as is the case for a perfect conductor:
\begin{equation}
    B_{\rm r} = 0, E_\theta = E_\phi = 0.
\end{equation}
To evolve the electromagnetic field over time, the relativistic MHD model solves the following conservation laws:
\begin{equation}
    \frac{\partial}{\partial t} \begin{bmatrix}D \\ \mathbf{m} \\ E_{\rm t}\\ \mathbf{B} \end{bmatrix} + \mathbf{\nabla} \cdot \begin{bmatrix}D\mathbf{v} \\ w_{\rm t} \gamma^2 \mathbf{v} \mathbf{v} - \mathbf{b} \mathbf{b} + I_3p \\ \mathbf{m} \\ \mathbf{v}\mathbf{B}-\mathbf{B}\mathbf{v} \end{bmatrix}^T = \begin{bmatrix}0 \\ \mathbf{f_{\rm g}} \\ \mathbf{v} \cdot \mathbf{f_{\rm g}}\\ \mathbf{0} \end{bmatrix}.
\end{equation}
Here, $\mathbf{m}$ is the momentum density, $D=\gamma \rho$ is the laboratory density with rest-mass density $\rho$, $p_{t}$ is the thermal pressure, $\mathbf{v}$ is the velocity of matter, $E_{\rm t}$ is the total energy, $\mathbf{b} = \mathbf{B}/\gamma+\gamma(\mathbf{v}\cdot\mathbf{B})\mathbf{v}$, and $\mathbf{f_{\rm g}} = \rho \gamma^2[\gamma^2\mathbf{v}(\mathbf{v} \cdot \mathbf{a})+\mathbf{a}]$ is a function of acceleration $\mathbf{a}$ \citep{Mignone_2007}.
An approximate Harten, Lax, and Van Leer Riemann Solver (\texttt{hll}) \citep{HLL} was used, as diffusive solvers are less likely to produce unwanted shocks in strongly supersonic or highly magnetized flows. Hyperbolic divergence cleaning \citep{div} was employed to ensure $\nabla \cdot B = 0$. In our setup, this method was found to evolve the magnetic field in unphysical ways if the simulation started with a stellar wind $v>0.81~c$, where $c$ is the speed of light. To address this problem, we implemented a workaround: the simulation begins at $v=0.808~c$ and the velocity of the pulsar wind is slowly increased to a final value of $0.985~c$. Beyond this value, the results again become unphysical, presumably due to the numerical methods being unsuited to such ultra-relativistic environments. The simulation parameters are provided in Table \ref{Konacki} for the known terrestrial planets, Table \ref{DiamondParam} for the ``diamond planet'' PSR J0636+5129 b, and Table \ref{TP} for a hypothetical terrestrial planet that could eventually be found.

\section{Computation of the radio emission}\label{comput}
\begin{figure*}
    \includegraphics[width=\columnwidth]{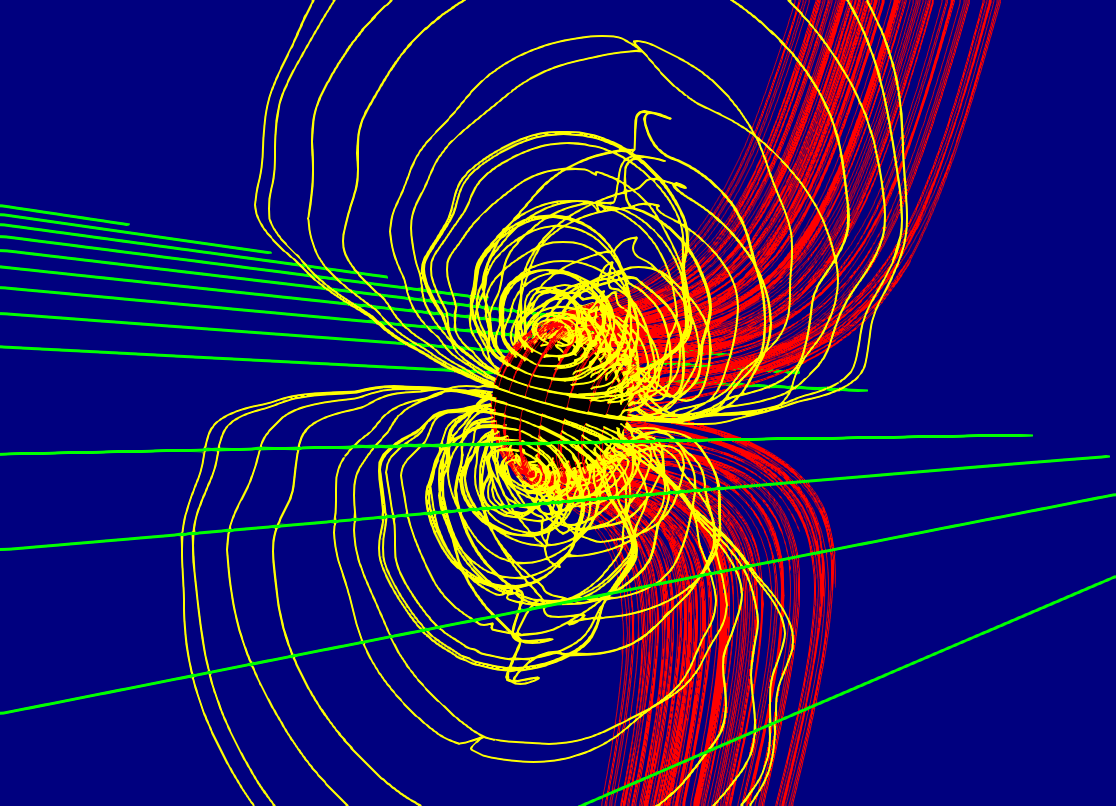}
    \includegraphics[width=0.9\columnwidth]{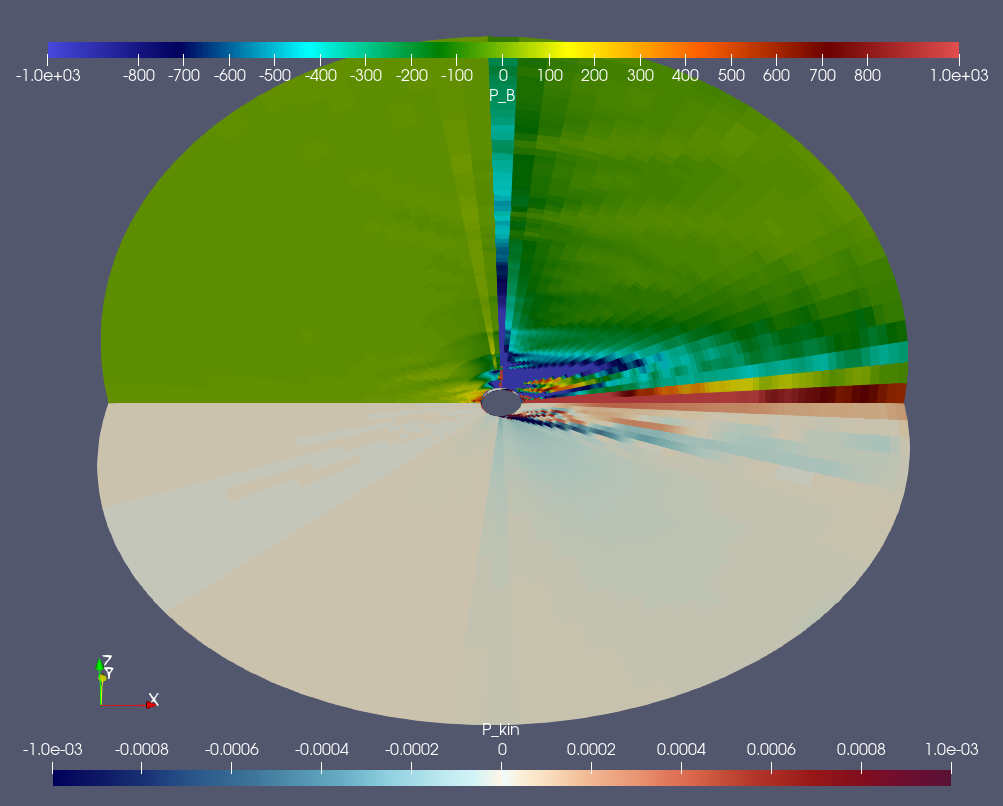}
    \caption{The resulting magnetic field and radio-emission with the pulsar wind velocity of $v=0.985~c$ in the case of PSR J0636+5129 b.
    In the left panel are shown the magnetic field lines (red) and electric currents (yellow) close to the planetary surface. The green lines represent the pulsar wind, directed from left to the right side. In the right panel, in the $xy$-plane is shown the emission due to the the dissipation of kinetic energy $\beta_{\rm kin}P_{\rm kin}$, and in the $xz$-plane is shown the emission due to the divergence of the Poynting flux, $\beta_{B}P_B$. $P_B$ dominates the spectrum by 7 orders of magnitude and shows multiple emission shells centered around the planet, which correspond to Alfv\'{e}n wings.}
    \label{Energies}
\end{figure*}

Fig.~\ref{Energies} shows the magnetic field lines bent around the planet, elongated at the nightside. The yellow lines  represent the currents, showing a dipolar structure that connects to the planetary surface.

\begin{table*}
\centering
\begin{tabular}{| c | c | c | c | c | c | c | c ||}   
\hline
Planet & Hyp. Planet  & Paper I & PSR B1257$+$12b & PSR B1257$+$12c & PSR B1257$+$12d & PSR J0636$+$5129b \\  
\hline
$\Phi_{\rm a}$(750) [Jy] & $6.87$ & $7.37\times10^{-3}$ & $2.02\times10^{-3}$ & $1.16\times10^{-2}$ & $5.81\times10^{-3}$ & $1.51\times10^2$ \\  
\hline
$\Phi_{\rm b}$(250) [Jy] & $6.19\times10^1$ & $6.64\times10^{-2}$ & $1.82\times10^{-2}$ & $1.04\times10^{-1}$ & $5.23\times10^{-2}$ &  $1.38\times10^3$\\  
\hline
$\Phi_{\rm c}$(100) [Jy] & $3.86\times10^2$ & $4.15\times10^{-1}$ & $1.14\times10^{-1}$ & $6.52\times10^{-1}$ & $3.27\times10^{-1}$ &  $8.48\times10^3$\\  
\hline
$\gamma$ & 5.798 & 2.0 & 5.798 & 5.798 & 5.798 & 5.798 \\  
\hline
$P_{\rm radio}$ [W] & $1.04\times10^{20}$ & $7.09 \times 10^{12}$ & $4.08\times10^{13}$ & $7.74\times10^{13}$ & $3.01\times10^{13}$ & $4.01\times10^{20}$ \\  
\hline
 $\Delta f$(80\%)[MHz] & $1.79\times10^2$ & $1.52\times10^{-3}$ & $2.41\times10^{-1}$ & $8.03\times10^{-2}$ & $6.35\times10^{-2}$ & $3.18\times10^{1}$\\  
\hline
$\Omega$ & 6.359 & 5.702 & 6.289 & 6.253 & 6.128 & 6.285\\  
\hline
$f_{\rm max, obs}$[MHz] & 4818.906 & 0.139  & 7.236 & 3.819 & 2.989 & 1788.237 \\  
\hline
LOFAR (750) & YES & NO & NO & NO & NO & YES \\  
\hline
MeerKAT(750) & YES & NO & NO & NO & NO & YES \\  
\hline
SKA (750) & YES & NO & NO & NO & NO & YES \\  
\hline
\end{tabular}
\caption{Expected intensity of the radio emission fluxes $\Phi_{\rm a}, \Phi_{\rm b}, \Phi_{\rm c}$ for planets at 750~pc, 250~pc and 100~pc from Earth, respectively. With the minimum sensitivities of LOFAR, MeerKAT and SKA of the order of 0.1, 0.01, and 0.001~mJy respectively, simulated pulsar planets would be detectable by each of them. The second column shows a hypothetical terrestrial planet which would be detectable and the third column is a reanalysis of the results of Paper I, now with emissions computed both from the nightside and the dayside. The total emission is roughly doubled compared to the previous result of $3.65\times10^{12}$~W. However, absorption in the Earth ionosphere still prevents detection with ground based telescopes. The 4th to 6th columns show that all confirmed terrestrial pulsar planets are not detectable using radio emission. The 7th column shows that the ``diamond planet'' PSR J0636+5129 b could be detected.}
    \label{Table}
    \end{table*}

The radio emission of pulsar planets is proportional to the dissipation of kinetic and magnetic energy \citep{Zarka07}. To compute the relativistic kinetic energy flux, we consider the conservation of the energy momentum tensor $T^{\mu\nu}$ in flat spacetime, with metric signature $(+,-,-,-)$:
\begin{equation}
    \partial_\mu T^{\mu \nu} = 0.
\end{equation}
In this case, we are interested in the dissipation of the energy density $E$, which means that we will focus on
\begin{equation}
    \partial_\mu T^{\mu 0} = \partial_0T^{00}-\partial_iT^{i0} = 0,
\end{equation}
\begin{equation}
    \partial_0 T^{00} = \frac{\partial E}{\partial t} = \partial_i T^{i0}.
\end{equation}
Approximating the pulsar wind as a perfect fluid with 4-velocity $u^\mu = \gamma (c,\mathbf{v})$, we can write:
\begin{equation}
    \frac{\partial E}{\partial t} = \partial_i\left[\left(\rho+ \frac{p}{c^2} \right)u^iu^0+pg^{i0}\right].
\end{equation}
In a flat spacetime, $g^{i0}=0$, so:
\begin{equation}
    \frac{\partial E}{\partial t} = \partial_i\left[\left(\rho+ \frac{p}{c^2} \right)u^iu^0\right].
\end{equation}
Rewriting the derivative:
\begin{equation}
    \frac{\partial E}{\partial t} = \gamma^2 c \vec{v}\cdot\vec{\nabla}\left(\rho+ \frac{p}{c^2} \right)+c\left(\rho+ \frac{p}{c^2} \right) \vec{\nabla}\cdot(\gamma^2\vec{v}).
\end{equation}
This means that the dissipated kinetic energy over some volume $V$ is equal to:
\begin{equation}
    P_{\rm kin} = \int_{\rm V}\left[ \gamma^2 c \vec{v}\cdot\vec{\nabla}\left(\rho+ \frac{p}{c^2} \right)+c\left(\rho+ \frac{p}{c^2} \right) \vec{\nabla}\cdot(\gamma^2\vec{v})\right] dV.
\end{equation}
In the right panel in Fig.~\ref{Energies}, the integral is not computed since the dissipated power is analyzed at specific points instead.

In addition to kinetic energy, there is also energy dissipated in the electromagnetic field. This is given by the divergence of the Poynting flux \citep{Varela_2022}, the computation of which does not change from the non-relativistic case:
\begin{equation}
    P_B = \int_V \vec{\nabla}\cdot\frac{(\vec{v} \wedge \vec{B}) \wedge \vec{B}}{\mu_0} dV.
\end{equation}

To estimate the emission, we can use the fact that the conversion efficiency of kinetic energy to emission is $\beta_{\rm kin} \sim 10^{-5}$, while the conversion efficiency of dissipated magnetic field energy to emission is $\beta_{B} \sim 2 \times 10^{-3}$ \citep{Zarka07}. Since these values are empirically measured, they will not change in the relativistic case. As shown in Fig.~\ref{Energies}, the magnetic emission dominates. Due to the formation of Alfv\'{e}n wings, the magnetic emission also shows a distinctive structure of multiple shells centered around the planet.

Throughout this work, dissipation rates are evaluated locally to identify emitting regions; total powers quoted are obtained by integrating over the emitting volume after applying cuts to exclude areas near coordinate singularities.

\section{Shape of strongly emitting regions}\label{shape}
There are various factors that must be considered when mapping the emission: in addition to the amount of emission, its origin and orientation must be accounted for.
\begin{figure*}[t]
    \includegraphics[width=0.82\columnwidth]{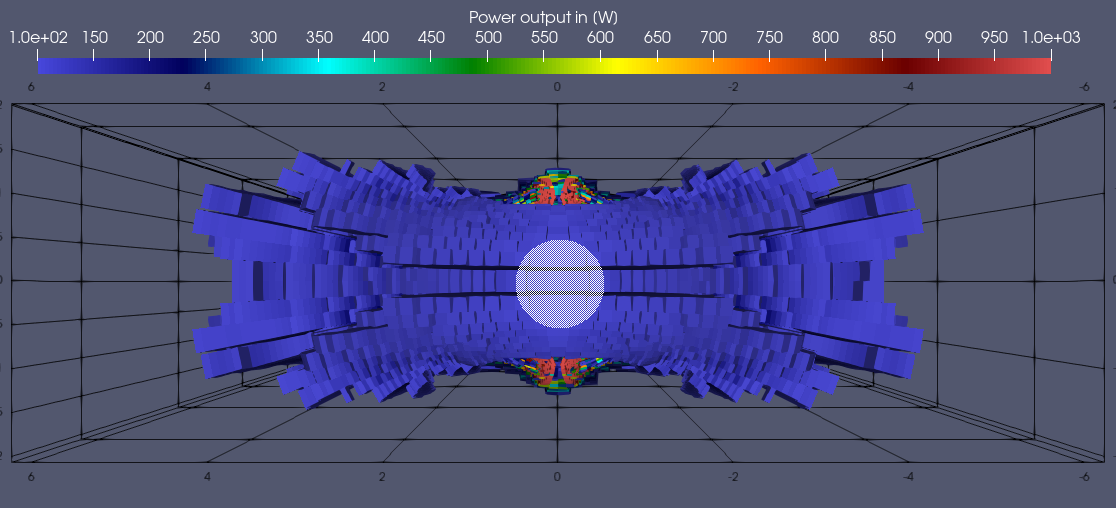}
    \includegraphics[width=1.17\columnwidth]{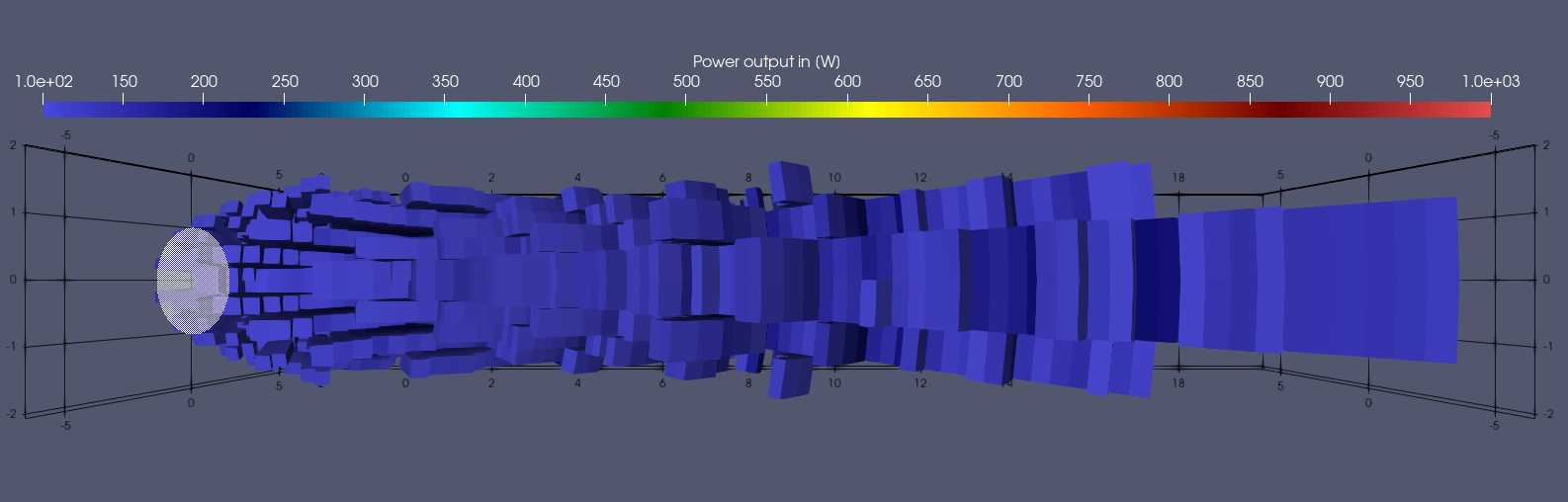}
    \caption{Plots illustrating the regions of strongest magnetic emission for the pulsar PSR J0636+5129 b in the conductive case. Shown are powers above $100$~W, with the grid in units of planetary radii. In the left panel, which shows the frontal view, the emitting region is aligned perpendicular to the magnetic field lines. The right panel displays the side view, where the region of strongest emission is positioned behind the planet. The planet itself is located within the white-shaded circular region.}
    \label{Shape}
\end{figure*}

In Paper I, only emissions on the dayside of the planet were considered. As shown in Fig.~\ref{Shape}, the nightside of the planet contributes with a strong emission, so it should be included as well. The region of strongest emission is aligned perpendicular to the external magnetic field. The two ``pillars'' of emission near the planetary poles are the artifact of the setup, so their contribution is excluded from the total emission.

With the locations of the emitting regions established, we need also to consider the direction of the emission. For this, we consider the Poynting vector within the strongest emitting regions, defined as areas where $P_{\rm radio,local} > 5.0 \times 10^{-18} \text{W} \times P_{\rm radio,tot}$. Since the planets all have different sizes and the emitting regions are asymmetric, this criterion ensures a consistent comparison across different simulations.

The emission strength depends on two competing factors: the local magnetic field magnitude and its angle relative to the pulsar wind. Downwind of the planet, the field is compressed and strongest directly behind it (`above' and `below' in Fig. \ref{Shape}). However, in these regions, the field lines run nearly parallel to the wind, which is a geometry that suppresses synchrotron-type emission. Conversely, at the sides of the planet (`left' and `right') the field is weaker, but its orientation is not parallel to the wind, which is the more favorable geometry for emission. The latter configuration dominates, making the lateral regions the primary emitters despite the lower field magnitudes.

\begin{figure*}[h]
    \includegraphics[width=\columnwidth]{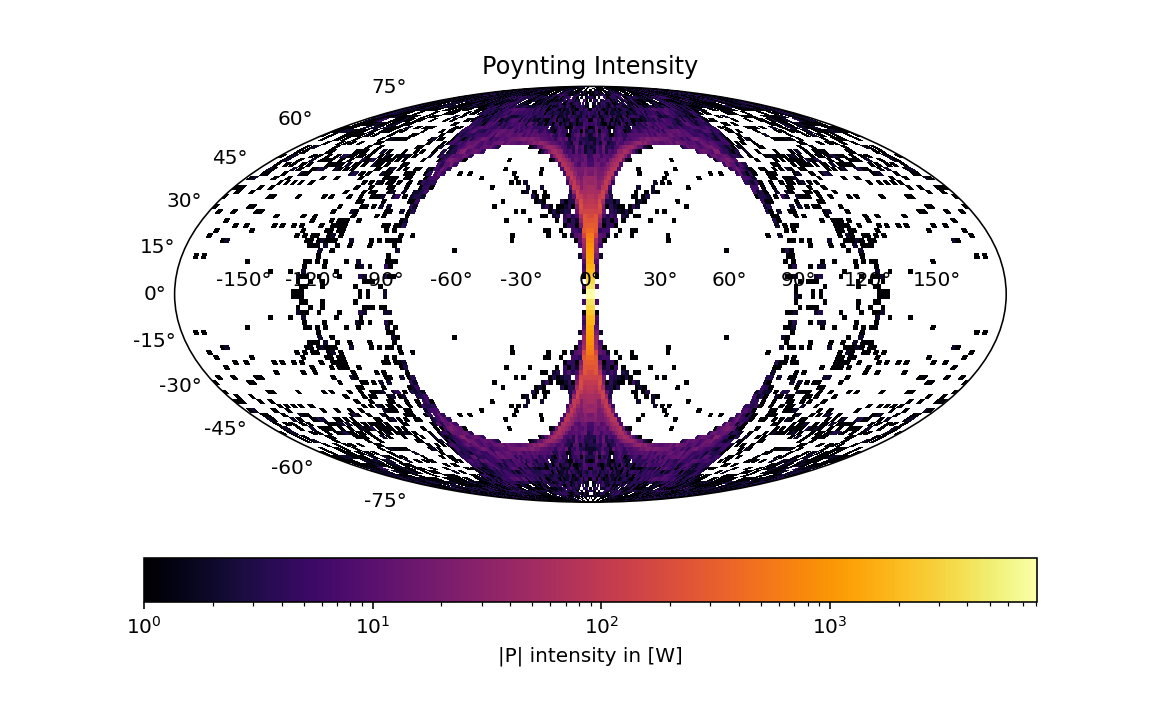}
    \includegraphics[width=\columnwidth]{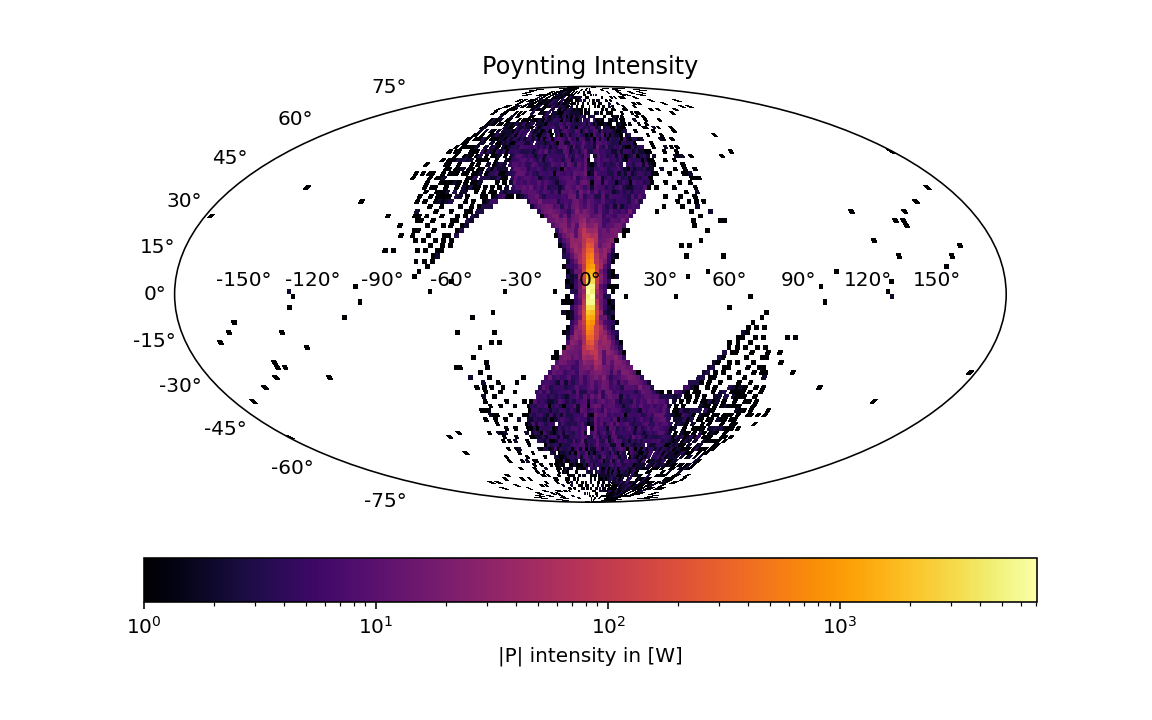}
    \caption{The Poynting vector was used to track the direction of emission for PSR J0636+5129 b in Mollweide projection. We consider cells that had a local Poynting vector magnitude above $P_{\rm radio,local} > 5.0 \times 10^{-18} \times P_{\rm radio,tot}$. The left panel shows the conductive case, for which the emission is hourglass shaped and spread concentrated near the equator, while it is further spread near the pole. The right panel shows the ferromagnetic case, where the emission distribution is S-shaped. Both have the strongest emission in a narrow line aligned with the magnetic field downwind of the planet.}
    \label{Poynting}
\end{figure*}

As shown in Fig.~\ref{Poynting}, the emission is not shaped like a perfect cone. To estimate the solid angle, we constructed the smallest surface that contains all Poynting vectors. This surface was then broken into small triangular elements formed between nearby vectors, and the total solid angle was approximated by summing their contributions.

The final remaining question is whether the planetary emission can be detected by ground-based observatories. In Fig.~\ref{Energies}, it was shown that magnetic emission dominates, whether or not the dissipated kinetic energy is computed while considering relativity (Paper I). For this reason, we set $P_{\rm radio} \approx \beta_{B}P_B$. The density of radio emission at a distance $d$ from the planet is \citep{Sergio}:
\begin{equation}
    \Phi = \gamma^{2}\frac{A_{c}}{\pi}\frac{P_{\text{radio}}}{d^{2}\Delta f}, 
\end{equation}
where $A_{c}$ is an anisotropy factor defined as $A_{c} = 4\pi/\Omega $, due to relativistic beaming compressing the emitting directions into a forward cone subtending a solid angle $\Omega$, and $\Delta f$ is the bandwidth of emissions.  Normalized to fiducial values, the relation takes approximately the following form:
\begin{equation}\label{l}
    \left(\frac{\Phi}{\text{Jy}}\right) = 10^{-24}\frac{4\pi}{\Omega}\left(\frac{\gamma}{10^5}\right)^2  \left(\frac{P_{\text{rad}}}{\text{W}}\right) \left(\frac{\text{Gpc}}{d}\right)^2 \left(\frac{\text{GHz}}{\Delta f}\right),
\end{equation}
with the solid angle of the beam being measured by plotting the Poynting vector over a circular surface surrounding the planet and integrating using a mesh of triangular simplices. Here $\gamma=5.795$ is the Lorentz factor in the pulsar wind. The frequency of the emitted light is estimated with the critical synchrotron emission, computed with the following expression \citep{Rybicki:847173}:
\begin{equation}
    f = \frac{3}{2}\gamma^2\frac{eB_{\rm sw}}{2 \pi m_{\rm e}}\sin{\alpha},
\end{equation}
where $e$ and $m_{\rm e}$ stand for the electron charge and mass, respectively. We can further simplify the computation by assuming isotropic pitch angles for the ensemble average \citep{Rybicki:847173}, where $\sin{\alpha}=\sqrt{2/3}$:
\begin{equation}
    f = \frac{3}{2}\gamma^2\frac{eB_{\rm sw}}{2 \pi m_e}\sqrt{\frac{2}{3}} \approx 3.43 \text{MHz}\gamma^2 \frac{B_{\rm sw}}{G}.
\end{equation}
This expression is valid in the frame of reference of the source. 

\begin{figure*}[h]
    \includegraphics[width=\columnwidth]{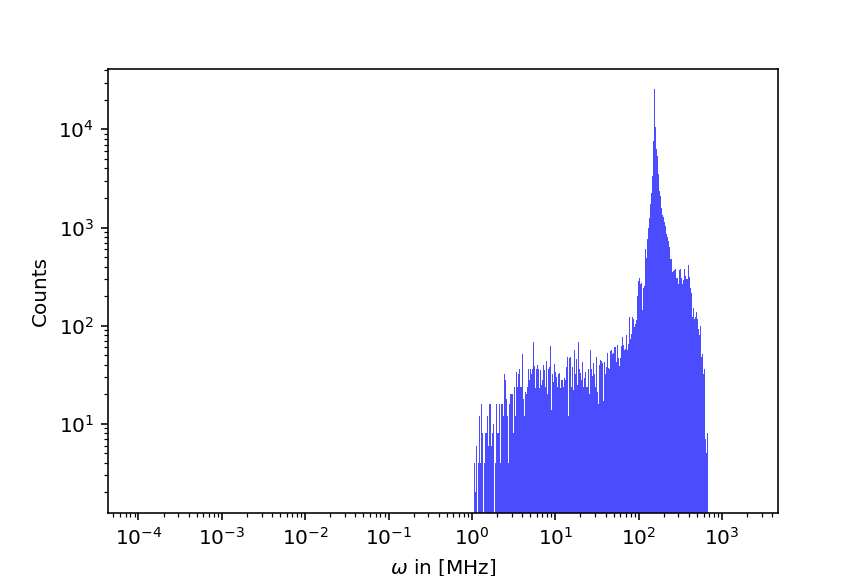}
   \includegraphics[width=\columnwidth]{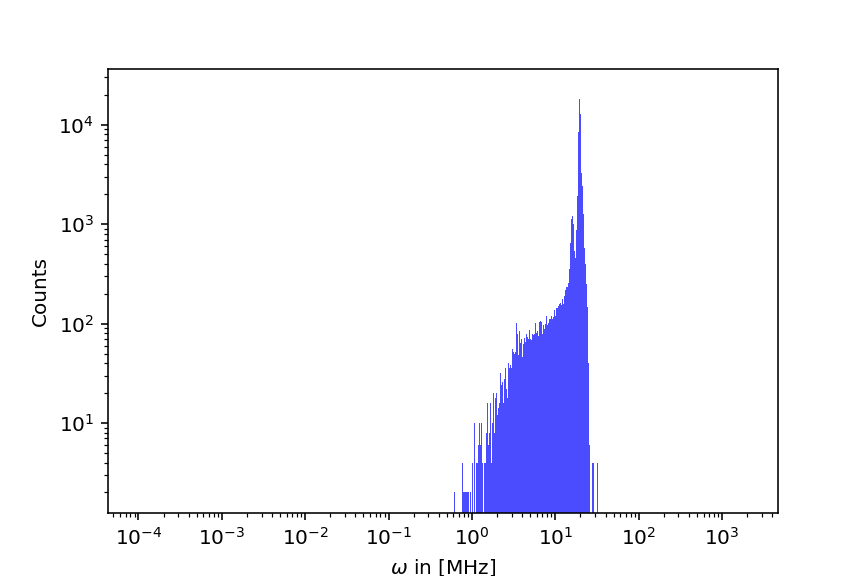}
    \caption{The frequency distribution for the conductive (left) and ferromagnetic (right) cases is shown in 1000 bin histograms. For both cases, we see a power law spectrum with a narrow peak and a high frequency cutoff. To isolate the frequency band, the 80-pecentile width was used. The conductive cases shows larger bandwidth but due to the used velocities being different it is unclear whether this is due to geometry or the velocity difference.}
    \label{Histograms}
\end{figure*}

To obtain the emission bandwidth, we apply this formula to the entire domain of integration. As shown in Fig.~\ref{Histograms}, the distribution is not symmetric, so we use the 80 percentile width.

While Equation~\ref{l} uses the bandwidth near the source, we must also verify that the radiation frequency is above the limit set by the Earth's ionosphere, which absorbs all frequencies below 10~MHz. To estimate this, we need the frequencies converted to the frame of the observer, $f_{\rm obs}$, which we find by using the Doppler effect formula:
\begin{equation}
    f_{\rm obs} = \frac{f}{\gamma(1-\beta\cos{\theta_{\rm obs}})},
\end{equation}
where $f$ is the frequency in the source frame, $\beta$ is the magnitude of the pulsar wind in units of $c$, and $\theta_{\rm obs}$ is the angle with respect to the line between the source and the observer. For head-on motion ($\theta_{\rm obs}=0$) and large speeds ($v\rightarrow c$), one obtains $[\gamma(1-\beta)]^{-1} \approx 2\gamma$. Then we can write:
\begin{equation}
    f_{\rm obs} = 2\gamma f.
    \label{i}
\end{equation}
To measure whether the radiation passes the ionosphere, we need to know if most of the signal is passing through it. To that end, we take the peak value of the frequency distribution and apply the Doppler shift.

The obtained radiation output is detailed in Table \ref{Table}. The second column details results for a hypothetical planet. The third column is a reanalysis of the results of Paper I, now with the emission originating in the nightside, recalculation of $\Phi_{\rm a}, \Phi_{\rm b}, \Phi_{\rm c}$ using equation \ref{l} and recalculation of $\Delta f_{\rm obs}$ using equation \ref{i}. The total emission is approximately doubled with inclusion of the nightside.

\section{Application to specific planets}\label{planets}

In this section, we present the results of the simulation for a specific planet, PSR J0636$+$5129 b \citep{fiore2023greenbanknorthcelestial}, a so-called ``diamond planet'', with a mass of 10$M_{\rm J}$ but a radius of only $0.74\pm0.136 R_J$. This means that it is likely the core of a gas giant, with its atmosphere torn off, because the radius is too small in comparison to its mass for a gas giant and the mass is too low for a white dwarf \citep{Kilic_2007}. Such cores are composed of metals, and thus the approximation as a conductor is appropriate. Since the radius and magnetic field of PSR J0636$+$5129 are unknown, the radius was assumed to be $10$~km and the magnetic field at its surface $10^8$~G, which are standard values for millisecond pulsars. The magnetic field at the location of the planet was then estimated using the split monopole approximation \citep{P_tri_2016}, using the fact that the pulsar period is known to be 2.87~ms and the semi-major axis of the orbit is $5.4\times10^{10}$~cm \citep{fiore2023greenbanknorthcelestial}. The simulation parameters are listed in Table \ref{DiamondParam}.
\begin{table}[h]
    \small
    \centering
    \begin{tabular}{|c|c|c|c|c|c|c|}
    \hline
    SWDens& SWSpeed &  SWTemp& SWMagField & PlanetRad \\
    $[{\rm g}/{\rm cm^3}]$ &  [${\rm cm}/{\rm s}$] & [$\rm K$] &[$\rm G$] & [$\rm cm$]  \\
\hline
 $3.1\times10^{-18}$ & $2.953\times10^{10}$ & $5.0\times10^8$ & $9.9$ & $5.0\times10^9$\\
\hline
    \end{tabular}
    \caption{Simulation parameters used for the ``diamond planet'' PSR J0636$+$5129 b.}
    \label{DiamondParam}
\end{table}
Is this planet observable? Its distance from us is estimated to be 500~pc \citep{fiore2023greenbanknorthcelestial}, and in Table~\ref{Table} we show that it would be visible even at 750~pc.

\subsection{Terrestrial planets}
\begin{figure*}[h]
    \includegraphics[width=0.33\textwidth]{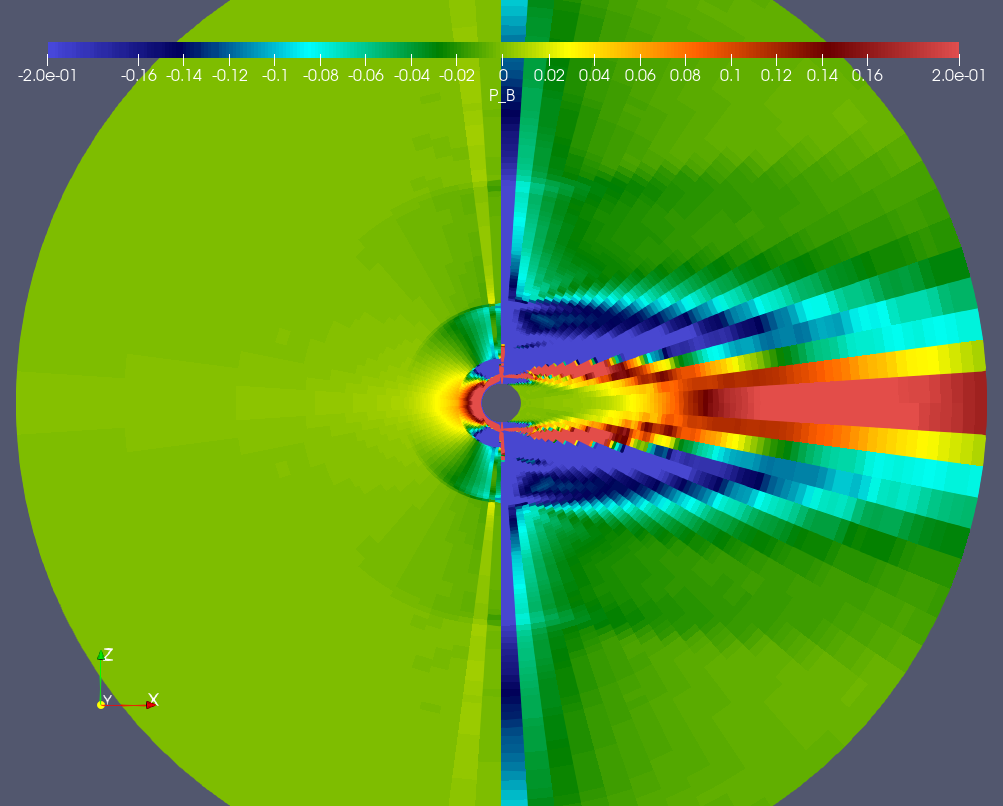}\
    \includegraphics[width=0.33\textwidth]{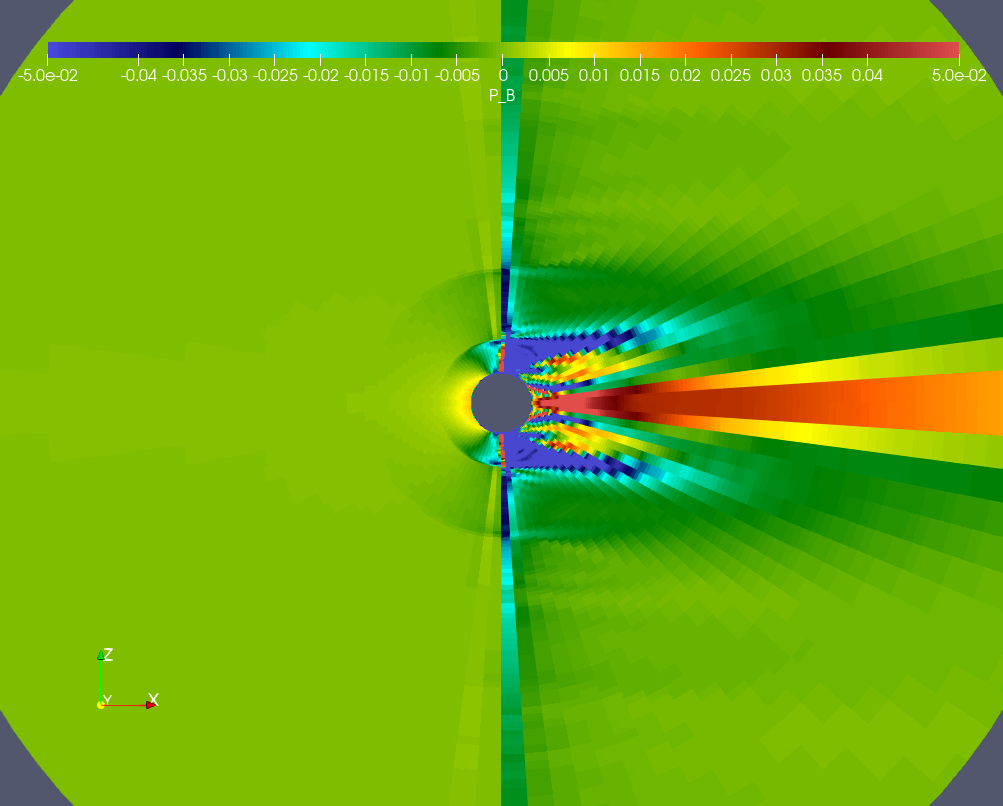}
    \includegraphics[width=0.33\textwidth]{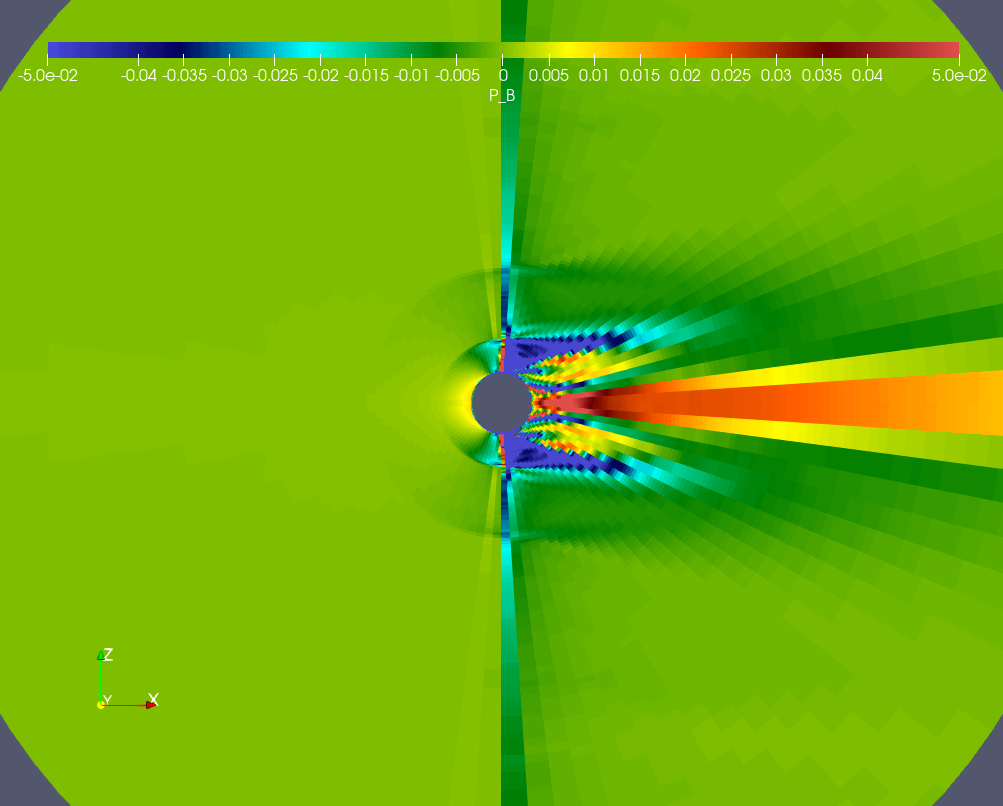}
    \caption{The emission in watts for the terrestrial planets PSR B1257$+$ 12 b (left), c(middle), d(right). Note that the scale for PSR B1257$+$ 12 b is different.}
    \label{Terr}
\end{figure*}
Here we focus on terrestrial planets, specifically the system in which the first planets were discovered. The pulsar with period of 6.219~ms houses three known planets: PSR B1257$+$ 12 b, c, and d \citep{Konacki_2003}. The rest of the parameters are unknown, so we consider a generic millisecond pulsar with a radius of 10~km and the magnetic field at its surface estimated with the split-monopole approximation as $10^8$~G. The parameters used in simulations and the planetary data are listed in Table~\ref{Konacki}.
\begin{table*}[h]
    \centering
    \begin{tabular}{|c|c|c|c|c|c|c|}
         \hline
    Planet& SWDens& SWSpeed &  SWTemp& SWMagField & PlanetRad & OrbRad \\
     & [$\frac{\rm g}{\rm cm^3}$] &  [$\frac{\rm cm}{\rm s}$] & [$\rm K$] &[$\rm G$] & [$\rm cm$] & [$\rm cm$] \\
    \hline
     PSR B1257+12 b & $3.1\times10^{-24}$ & $2.953\times10^{10}$ & $5.0\times10^8$ & $4.1\times10^{-2}$ & $2.0\times10^8$ & $2.8\times10^{12}$\\
    \hline
     PSR B1257+12 c & $3.1\times10^{-24}$ & $2.953\times10^{10}$ & $5.0\times10^8$ & $2.1\times10^{-2}$ & $1.0\times10^9$ & $5.4\times10^{12}$\\
    \hline
     PSR B1257+12 d & $3.1\times10^{-24}$ & $2.953\times10^{10}$ & $5.0\times10^8$ & $1.6\times10^{-2}$ & $7.5\times10^8$ & $6.9\times10^{12}$\\
    \hline
    \end{tabular}
    \caption{Wolszczan's terrestrial planets.}
    \label{Konacki}
\end{table*}
From the Table~\ref{Table} we see that we can not detect these planets. The low magnetic fields at the planetary location limits the emission bandwidth too much to allow observation. Even when strong enough, the signal would be absorbed in the Earth ionosphere. 
\begin{table*}
    \centering
    \begin{tabular}{|c|c|c|c|c|c|c|c|c|c|}
        \hline
        SWDensity& SWSpeed &  SWTemp& NSRadius & NSPeriod  & NSField & Orbital Radius  & SWMagField & PlanetRad \\
        $[\frac{\rm g}{\rm cm^3}]$ &  [$\frac{\rm cm}{\rm s}$] & [$\rm K$] & [$\rm cm$] &  [$\rm ms$] & [$\rm G$] &  [$\rm cm$] &  [$\rm G$] & [$\rm cm$] \\
        \hline
         $3.1\times10^{-17}$ & $2.953\times10^{10}$ & $5.0\times10^8$ & $10^6$ & $1$ & $10^8$ &  $1.2\times10^{12}$ & $3.6$ & $5.0\times10^8$\\
        \hline
         $3.1\times10^{-17}$ & $2.953\times10^{10}$ & $5.0\times10^8$ & $10^6$ & $6$ & $10^8$ &  $3.4\times10^{10}$ & $3.6$ & $5.0\times10^8$\\
        \hline
    \end{tabular}
    \caption{Hypothetical planet data}
    \label{TP}
\end{table*}
\subsection{Hypothetical planets}\label{Theo}
The three planets in the system PSR B1257+12 found by \cite{Wolszczan1992} are currently the only known terrestrial pulsar planets. Since we find that their radio emission is not detectable for the surface instruments, the next question is what parameters would allow for an observation of terrestrial planet?

The main limiting factor is the magnetic field at the planetary location. It affects the emission in two ways:\\
1. The emission peak, which needs to be above a threshold of 10~MHz, is proportional with the magnetic field strength near the planet.\\
2. Radiated power is dominated by dissipation of the magnetic field, which means that a higher field leads to stronger emission.

To satisfy this, we set the external field at the planetary location to 3.6~G, since this guarantees that the threshold will be above the limit set by the ionosphere, the host star is assumed to be a generic millisecond pulsar and the planet radius is set to 5000~km. We then estimate the planetary parameters with the split-monopole approximation.

\begin{table*}[h]
\centering
\small
\begin{tabular}{| c | c | c | c | c | c | c | c |}   
\hline
Planet & Hyp. Planet &  Paper I & PSR B1257$+$12 b & PSR B1257$+$12 c & PSR B1257$+$12 d & PSR J0636$+$5129 b \\  
\hline
$\Phi_{\rm a}$(750) [Jy] & $2.01\times10^1$ & $1.31$ & $2.48\times10^{-3}$ & $1.20\times10^{-2}$ & $1.15\times10^{-2}$ & $4.04\times10^1$ \\  
\hline
$\Phi_{\rm b}$(250) [Jy] & $1.81\times10^2$ & $1.18\times10^{1}$ & $2.23\times10^{-2}$ & $1.08\times10^{-1}$ & $1.03\times10^{-1}$ &  $3.64\times10^2$\\  
\hline
$\Phi_{\rm c}$(100) [Jy] & $1.13\times10^3$ & $7.36\times10^{1}$ & $1.40\times10^{-1}$ & $6.73\times10^{-1}$ & $6.46\times10^{-1}$ &  $2.27\times10^3$\\  
\hline
$\gamma$ & 2.799 & 2.0 &  2.799 & 2.799 & 2.799 & 2.799 \\  
\hline
$P_{\rm radio}$ [W] & $5.66\times10^{19}$ & $4.14\times10^{15}$& $4.66\times10^{12}$ & $4.10\times10^{13}$ & $1.38\times10^{13}$ & $2.49\times10^{19}$ \\  
\hline
$\Delta f$(80\%) [MHz] & $8.827$ & $6.442\times10^{-3}$ & $5.152\times10^{-3}$ & $9.587\times10^{-3}$ & $3.361\times10^{-3}$ & $1.737$\\  
\hline
$\Omega$ & 5.585 & 4.426 & 6.415 & 6.257 & 6.260 & 6.206\\  
\hline
$f_{\rm max, obs}$[MHz] & 527.367 & 0.558 & 0.446 & 0.873 & 0.179 & 109.206\\ 
\hline
LOFAR (750) & YES & NO & NO & NO & NO & YES \\  
\hline
MeerKAT (750) & YES & NO & NO & NO & NO & YES \\  
\hline
SKA (750) & YES & NO & NO & NO & NO & YES \\  
\hline
\end{tabular}
\caption{This table lists the same parameters as Table \ref{Table}, but for the ferromagnetic case. In general, the conclusions remain the same as for the conductive case, but due to instabilities near the poles of the model the velocity could only be raised up to $0.93~c$.}
    \label{Ferro}
    \end{table*}
Table \ref{TP} shows that the pulsar period strongly influences the field strength near the planet. By increasing it from 6~ms to 1~ms, the orbital radius required to reach a magnetic field of 3.6~G has lowered by 2 orders of magnitude. This means that pulsars with shorter periods are better suited for the radio emission detection method, since planets do not need to be so close to the pulsar.

The results of the simulation are presented in the second column of Table~\ref{Table} and show that both the magnitude of the emission and its bandwidth should be more than enough to detect such a planet. The main limiting factor is the external magnetic field.

\subsection{Ferromagnetic planet}

In addition to the conductive case, following the Paper I, we also considered a ferromagnetic setup. The boundary conditions in this case are $B_{\rm r} = 0$, $B_\phi \rightarrow -B_\theta$, $B_\theta \rightarrow B_\phi$. The strong self-coupling of the magnetic field means that this setup was particularly unstable; to run it the velocity had to be lowered to $2.88\times10^{10}~\rm cm~s^{-1}$ (0.93~c) and a pressure maximum of $3\times10^7~\rm g~cm^{-3}$ was introduced.

In general, the ferromagnetic case produces results similar to those of the conductive case. As shown in Table \ref{Ferro}, the same planets are predicted to be visible as in the conductive case. The reason for the sometimes lower total energy than in the previous simulation in Paper I is probably the cap on pressure, resulting in energy being lost in the affected cells.

\section{Observational prospects}\label{obs}
The predictions from our simulations impose strong constraints on systems that can potentially serve as observational evidence. The intrinsic population for such systems is described by a luminosity function defined as a number density per unit luminosity,
\begin{equation}
\phi(L) \equiv \frac{dN}{dL \, dV}.
\end{equation}
These include the fraction of galactic population of pulsars hosting suitable companions (planets or asteroids), then the fraction of such systems in which the wind-companion interaction produces an emission region which is set by the pulsar wind magnetic field, its Lorentz factor, and the orbital elements. In addition, geometric selection effects such as the beam opening angle, the fraction of observers intersecting the emission beam and temporal effects, such as the duty cycle of the emission, further modulate the effective observable population. Only systems satisfying these physical conditions contribute to the emitting population. An additional constraint arises from propagation through the terrestrial ionosphere: only emission above the cutoff frequency ($\sim 10\,\mathrm{MHz}$) is observable from the ground.

The observable flux density is given by
\begin{equation}
S = \frac{L}{4\pi d^2},
\end{equation}
where $d$ is the distance to the source.

The cumulative source counts above a given sensitivity threshold $S$ are then
\begin{equation}
N(>S) = \int_{\mathrm{Galaxy}} dV \int_{L > 4\pi d^2 S} \phi(L)\, dL.
\end{equation}

For a survey with sensitivity limit $S_{\mathrm{lim}}$, sky coverage $\Omega$, and observing time $T$, the expected number of detections is
\begin{equation}
N_{\mathrm{det}} = T \, \Omega \, N(>S_{\mathrm{lim}}).
\end{equation}

Observability is determined by whether $N_{\mathrm{det}}$ is statistically significant. Non-detection is expected if $N_{\mathrm{det}} \ll 1$, while $N_{\mathrm{det}} \gg 1$ with no detections would imply that the assumed physical conditions or population fractions are incorrect.

Since different physical models would predict different observable populations, model comparison is best performed in terms of $N(>S)$. Each theoretical model produces a distinct curve $N_i(>S)$ in the $N-S$ space, obtained from fitting $N(>S)$ for different survey sensitivities.

With the described constraints, a blind broadband survey focused on fast-rotating millisecond pulsars seems to be the best observational strategy. It would accumulate statistical significance on many potential hosts. A non-detection from such a survey would be an additional information in search for the explanation of the pulsar planets scarcity. 

\section{Conclusions}\label{Concl}

Our special relativistic simulations of the magnetospheric interaction between a pulsar wind and orbiting planets demonstrate that radio emission is a viable detection method for pulsar planets. The emission is dominated by magnetic energy dissipation, and the Alfvén wing structure produces a signal strong enough to be detectable at interstellar distances with the current operating telescopes.

The known system PSR J0636+5129 b serves as a proof of concept: its predicted flux exceeds the sensitivity thresholds of LOFAR, MeerKAT, and SKA, confirming that the physical conditions required for detection are realistic. However, targeting specific systems requires prior knowledge of the orbital geometry and magnetic field orientation to determine whether the emission beam intersects the line of sight, which is a significant observational constraint.

For this reason, the most promising observational strategy is a blind broadband survey focused on fast-rotating millisecond pulsars. Such a survey avoids the need to model the beam orientation for individual systems, naturally accumulates statistics across many potential hosts, and preferentially selects for the systems most likely to produce detectable emission — since shorter pulsar periods yield stronger magnetic fields at the planetary orbit. A non-detection in such a survey would itself be a meaningful constraint on the pulsar planet population.

\section*{Acknowledgements}
M\v{C} acknowledges the Czech Science Foundation (GAČR) grant No. 21-06825X. PM work was partially funded by National Science Centre, Poland, grant no. 2023/49/B/ST9/01783. H.K. Vedantham of ASTRON, Netherlands is thanked for the observational insights. We thank the PLUTO team for the possibility to use the code.


\bibliographystyle{aasjournal}
\bibliography{nlibrary}

\end{document}